\documentclass[%
superscriptaddress,
aps,
prd,
rsi,%
amsmath,amssymb,
reprint,%
floatfix,%
a4paper]{revtex4-2}

\usepackage{graphicx}
\usepackage{dcolumn}
\usepackage[dvipsnames]{xcolor}
\usepackage{float}

\usepackage{mathrsfs}
\usepackage{mathtools}
\usepackage{relsize}
\usepackage{datenumber}
\usepackage[utf8]{inputenc}
\usepackage{cancel}
\usepackage{comment}
\usepackage{soul}

\usepackage{xspace}
\usepackage{amsmath}

\usepackage{multirow}
\usepackage{array}

\usepackage{lipsum}
\usepackage{subfigure}
\usepackage[justification=raggedright]{caption}

\usepackage{currfile}
\usepackage[yyyymmdd,hhmmss]{datetime}

\begin{document}


\author{Mohit Chaudhary}
\affiliation{ 
Aix-Marseille Univ., CNRS, CINAM, Marseille, France
}%
\affiliation{ 
European Theoretical Spectroscopy Facility (ETSF), Web: http://www.etsf.eu
}%

\author{Hans-Christian Weissker}
\email{hans-christian.weissker@univ-amu.fr}
\affiliation{ 
Aix-Marseille Univ., CNRS, CINAM, Marseille, France
}%
\affiliation{ 
European Theoretical Spectroscopy Facility (ETSF), Web: http://www.etsf.eu
}%

\author{Daniele Toffoli}
\affiliation{ 
Department of Chemical and Pharmaceutical Sciences, University of Trieste, Trieste 34127, Italy
}%
\affiliation{ 
CNR - Istituto Officina Dei Materiali (IOM), Trieste 34149, Italy
}%

\author{Mauro Stener}
\affiliation{ 
Department of Chemical and Pharmaceutical Sciences, University of Trieste, Trieste 34127, Italy
}%
\affiliation{ 
CNR - Istituto Officina Dei Materiali (IOM), Trieste 34149, Italy
}%

\author{Victor Despré}%
\affiliation{ 
Université Claude Bernard Lyon 1, CNRS, Institut Lumière Matière, UMR5306, F-69100 Villeurbanne, France
}%

\author{Franck Rabilloud}%
\email{franck.rabilloud@univ-lyon1.fr}
\affiliation{ 
Université Claude Bernard Lyon 1, CNRS, Institut Lumière Matière, UMR5306, F-69100 Villeurbanne, France
}%

\author{Jean Lerm\'e}%
\affiliation{ 
Université Claude Bernard Lyon 1, CNRS, Institut Lumière Matière, UMR5306, F-69100 Villeurbanne, France
}%

\author{Rajarshi Sinha-Roy}%
\email{rajarshi.sinha-roy@univ-lyon1.fr}
\affiliation{ 
European Theoretical Spectroscopy Facility (ETSF), Web: http://www.etsf.eu
}%
\affiliation{ 
Université Claude Bernard Lyon 1, CNRS, Institut Lumière Matière, UMR5306, F-69100 Villeurbanne, France
}%


\title{Surface Plasmons in the Continuum}

\date{\today}


\begin{abstract}
\noindent
\textbf{Abstract:}
The interest to foster plasmonic applications at energies in the ultra-violet, has escalated research initiatives in clusters of unconventional plasmonic materials like aluminum and indium, for which the surface-plasmon resonance appears above the ionization potential. Naturally, the quantum mechanical description calls for the incorporation of the ionization process, thereby making the \textit{ab initio} calculations challenging. We present a robust approach within the time-evolution formalism of the time-dependent density-functional theory to calculate surface plasmon resonance in the continuum of metal clusters. Using the much studied Al$_{13}^-$ as a system of reference, we show that accurate description of the continuum and of the ionization of the cluster allow to capture a broad surface-plasmon in the UV. Application of this approach in aluminum clusters has given the size-dependent evolution from discrete spectral features in Al$_{6}$ to the surface-plasmon in larger clusters in the deep ultra-violet. 
\end{abstract}


\maketitle

While traditional plasmonic materials, like gold or silver, are active in the visible or near-ultra-violet domains, the $sp$ metals like aluminum and indium are expected to have surface-plasmon resonances (SPRs) at far higher energies~\cite{Ehrenreich1963,Lemonnier1969,Blaber2007}, promising the realization of ultra-violet (UV) plasmonics~\cite{Ekinci2008,Knight2014,McKenna2025}.
Envisaged applications in the UV include surface-enhanced Raman scattering~\cite{Das2017}, 
high-energy photon harvesting through hot-carrier generation~\cite{Gong2017,Dong2025}, 
fluorescence enhancement~\cite{Priya2022} for bio-molecule detection~\cite{Barulin2019}, magnetoplasmonics~\cite{Matsuda2024}, high-responsivity photodetectors~\cite{Dubey2020}, and catalysis~\cite{Zhou2016,Robatjazi2017}.

Experimentally, relatively high plasmon energies have been obtained for  Al and In nanoparticles (NPs)  deposited in matrices~\cite{Ross-Schatz-2014,Cottancin2014}. Due to the formation of oxide layers and the interaction with the respective matrices, the SPRs are far lower in energy than they would be for free, clean NPs~\cite{Knight2014,Ross-Schatz-2014}. However, measuring the optical response of such free, clean aluminum or indium NPs is very challenging due to their strong tendency to undergo rapid oxidation, but also due to the difficulty of performing measurements in the UV. Efforts are made in thin-film deposition~\cite{Wang2021} to improve the UV plasmonic activity of Al. However, to the best of our knowledge, experimental spectra of free, clean NPs of Al or In are still missing in the literature.

From a theoretical perspective, describing plasmons at UV energies is challenging when the plasmon appears above the ionization potential (IP), in which case the effects of the ionization continuum must be taken into account. In real-time approaches, where the system's time evolution is explicitly computed, it is necessary to account for their coupling to the continuum, which enables these bound states to undergo autoionization. These aspects are rigorously taken into account in approaches based on the time-dependent Schrödinger equation in the atomic, molecular, and optical physics communities~\cite{Krause1992,Grobe1999,Chelkowski1998,Scrinzi2010}, but also using time-dependent density-functional theory (TDDFT) for energy-resolved photoemission~\cite{DeGiovannini2012,bachau-01,luccese-86}, photoionization~\cite{Calvayrac2000,Neidel2013,Wopperer2017,jcp_1.1937367}, absorption~\cite{Vincendon2018} in clusters, and charge migration in molecules~\cite{GuiotDuDoignon2025}. However, they have not yet been considered for describing surface-plasmon resonances above the IP in the UV domain.

Incidentally, small aluminum clusters appear appealing for calculations~\cite{Martinez1994,Akola1998_IPs,Deshpande2003,Shinde2014,Debnath2015,Akola1999_PES,Dolgounitcheva1999,Gobel2021,Casanova2016,Pandeya2021,Gao2025} because the number of valence electrons that need to be taken into account explicitly is low -- two $s$ and one $p$ electrons per atom. The anion Al$_{13}^{-}$  exhibits a high relative stability in mass spectra~\cite{Luo2014} and in the presence of oxygen~\cite{Leskiw2000}. This is due to combined shell closures as it adopts an icosahedral geometry and a closed-shell electronic structure with 40 valence electrons~\cite{Sengupta2016}. Consequently, this anion has been used in several theoretical studies~\cite{Akola1999_PES,Dolgounitcheva1999,Casanova2016,Gobel2021,Pandeya2021} as a system that is presumably simple, expected to show a clear SPR above the IP~\cite{Pandeya2021} in the UV, and that can be easily coupled with molecules to study the energy transfer between plasmons and molecular resonances~\cite{Gobel2021,Pandeya2021}.

However, the TDDFT-calculated spectra of Al$_{13}$~\cite{Deshpande2003,Debnath2015} and its anion Al$_{13}^-$~\cite{Pandeya2021,Casanova2016} in the literature show different features depending on the
density functional used, and the spatial representation (quality of atomic basis set or of grid simulation) [cf., Figure S1 in the Supplemental Material]. 
While they are consistent with the general Mie-theory based prediction~\cite{Blaber2007}: surface plasmon in Al clusters should occur in the UV, the large discrepancy in the literature poses questions on the reliability of TDDFT to describe SPR above the IP.

In the present work, we calculate the UV plasmons of aluminum clusters of sizes upto $\approx 2$ nm.  In contrast to previous works, we present a method to calculate SPRs above the IP by taking into account the coupling to the continuum and, consequently, the ionization process. To this end, we demonstrate that grid-based real-space real-time (RT) TDDFT with absorbing boundary conditions can properly describe the ionization. The much-studied anion Al$_{13}^-$ is considered as a system of reference for the validation of the method before we apply it to larger Al clusters. In addition, we present a linear-response TDDFT approach based on B-splines. This approach provides an accurate description of scattering and photoemission processes but has not previously been applied to plasmons in the continuum. Comparison with real-time TDDFT results shows that the problem of ionization is general and can be treated in different ways in TDDFT in order to describe plasmons in the continuum.
~\\

\begin{figure}
\includegraphics[width=.48\textwidth]{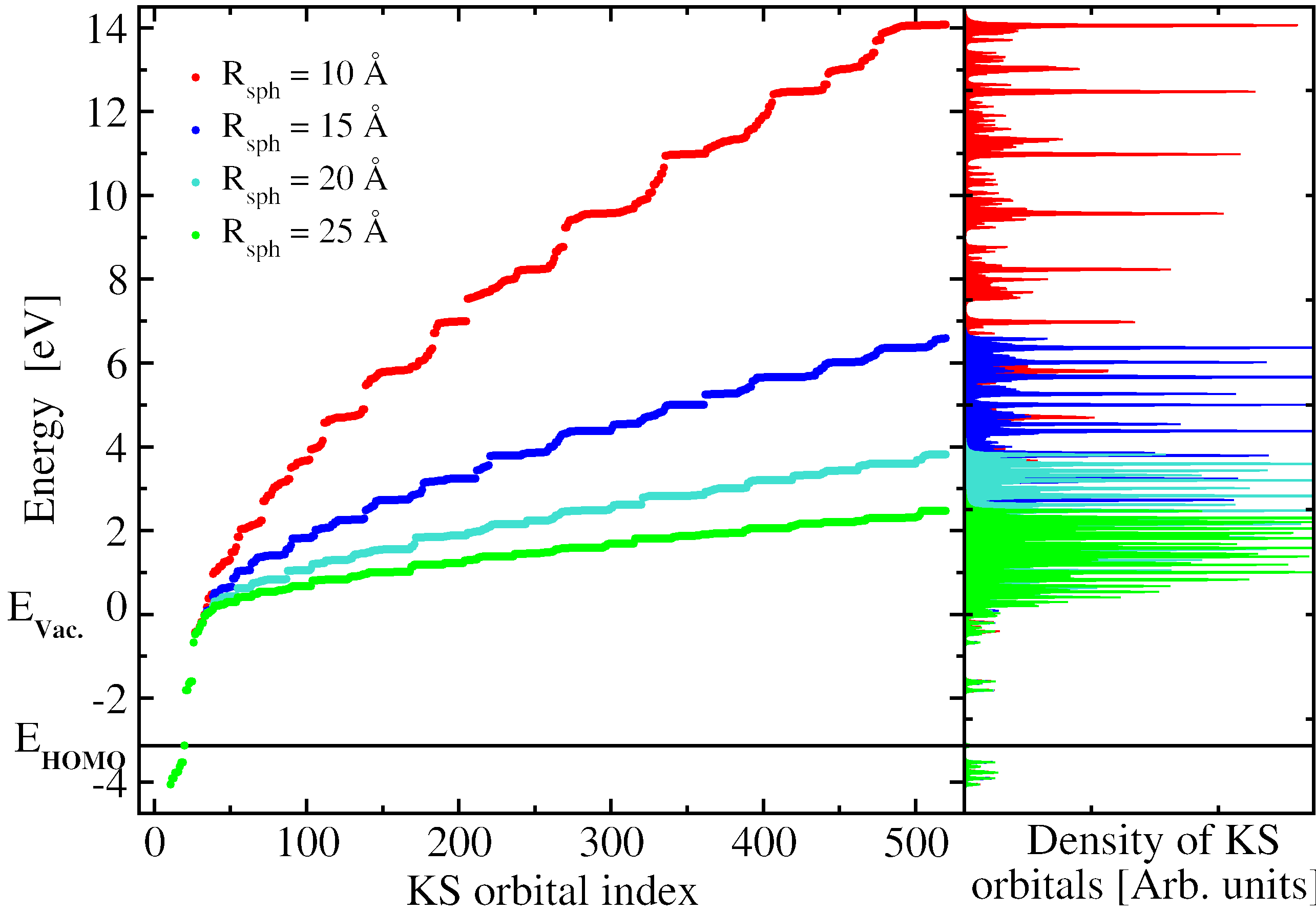}
\caption{
Energy of the first 500 Kohn-Sham orbitals and the corresponding density of states for different sizes of the spherical computation domain for Al$_{13}^{-}$. 
In all cases, the same number of orbitals are shown; they are denser for larger $R_{\rm sph}$ but cover an accordingly smaller energy range. 
}
\label{fig:eigenvalsNdos}
\end{figure}

\begin{figure}
\includegraphics[width=.485\textwidth]{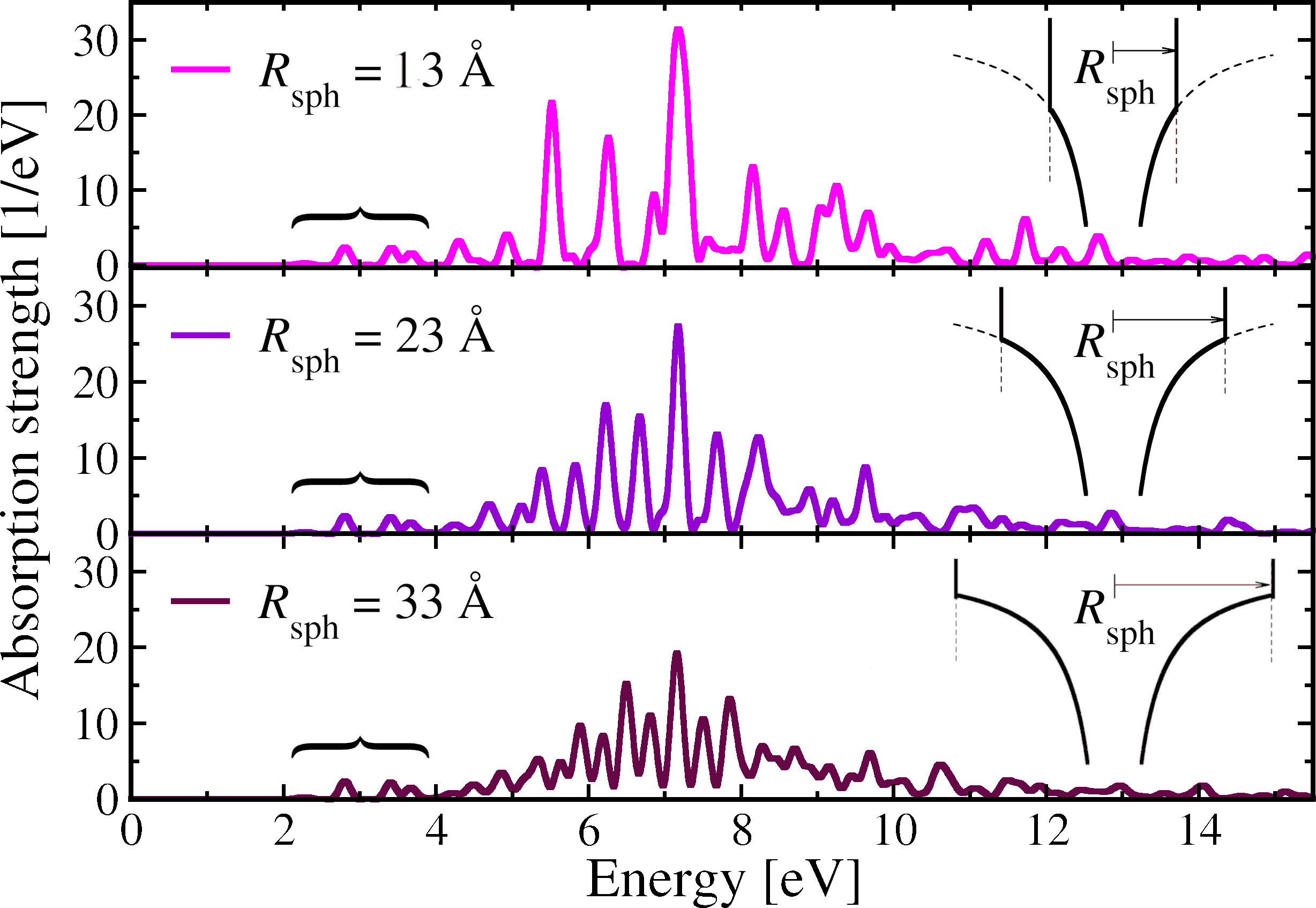}
\caption{
Absorption spectra of Al$_{13}^-$ calculated in RT-TDDFT, for different radii $R_{\rm sph}$ of the spherical simulation domain. The low-energy part of the spectrum that corresponds to bound-bound transitions and does not change with the domain size is indicated by the curly brackets. The inset shows a schematic view of the potential of the cluster along with the additional confining potential due to the finite spherical domain.
}
\label{fig:withNwoAbsBound}
\end{figure}

For calculating absorption using TDDFT, the numerical solution representing the electronic structure and the dynamics of the  system, i.e., the static and time-dependent Kohn-Sham orbitals, are obtained within some calculation domain, which can be problematic for the description of the continuum.
Indeed, the finiteness of the simulation box forces the solution to be zero at the box boundary, and thus acts as an additional confinement potential leading to an artificial discretization of the continuum states.
However, in grid-based approaches, this discretized representation of the continuum can be systematically improved by increasing the size of the simulation box and decreasing the spacing between the spatial grid points.
For this very reason we perform grid-based RT-TDDFT calculations using the code octopus~\cite{octopusTancogne-Dejean2020}. 
A spacing of $0.18\,\rm\AA$ is used to construct the grid in a spherical simulation domain of radius $R_{\rm sph}$. Atom-centered norm-conserving Troullier-Martins pseudopotential~\cite{Troullier1991} is used to describe the ionic cores of the atoms in the cluster. The adiabatic gradient-corrected PBE~\cite{Perdew1996GGAPBE} approximation is used for the exchange-correlation contributions in the static and time-dependent calculations. In addition, a self-interaction correction scheme based on the average-density~\cite{Legrand2002adsic} is also employed for the proper description of the asymptotic behavior of the Kohn-Sham (KS) potential, which is important for the accurate description of the IP and the surface-plasmon energy~\cite{Pacheco1992}.

In Figure~\ref{fig:eigenvalsNdos} we show the dependence of the ground state electronic structure of Al$_{13}^-$ on $R_{\rm sph}$: 
increasing the size of the simulation domain increases the density of the KS orbitals above the vacuum level, thereby suggesting an improvement of the description of the continuum, potentially leading to a better description of the coupling of the eigenstates of the system with the continuum. Figure~\ref{fig:withNwoAbsBound} shows RT-TDDFT absorption spectra of Al$_{13}^-$ calculated in real-space grids with radii ranging from $R_{\rm sph}=13\,\rm\AA$ to $33\,\rm\AA$, and for an evolution time of 24 fs. The low-energy part of the spectra ($< 4\,\rm eV$) does not depend on the domain size. This part corresponds to transitions to bound excited states ($E_{_\mathrm{HOMO}} = -3.13\,\rm eV$) that are not impacted by the continuum. However, at higher energies, increasing the radius of the simulation domain makes more discrete features appear in the spectrum. These features resembling interference patterns continue to change with increasing domain size and do not converge. Therefore, although enlarging the simulation domain improves the density of the discretized continuum states (taken into account implicitly in the RT formalism), the description of the continuum and the resulting spectra remains unsatisfactory.

A possible solution becomes clear when the physical process of ionization is considered within the RT approach. As transitions to eigenstates above the IP lead to subsequent autoionization of the system, it becomes crucial to take the quick disappearance of the emitted electrons from the system into account. The vast majority of the electrons in the continuum possess enough kinetic energy to reach the domain boundary in a very short time, much shorter than the period over which the time evolution is carried out. They will eventually get reflected at the boundary and interfere in the evolution of the system, in particular with the dipole moment needed for the calculation of the optical spectrum. To prevent this problematic behavior, we would need to perform the calculations in an infinite simulation domain. This, however, is clearly not feasible.

Therefore, in order to prevent this unwanted reflection, we employ absorbing boundary conditions~\cite{Krause1992,DeGiovannini2015} in the RT-TDDFT. 
A complex absorbing potential (CAP) as detailed in Ref.~\onlinecite{DeGiovannini2015} and as implemented in the code octopus~\cite{octopusTancogne-Dejean2020} is used to model the absorbing boundary (AB). 
The CAP is modeled as a sine-squared function within an absorbing region padded at the boundary of the simulation domain. 
The amplitude $\eta$ of this function gives the height of the CAP. 
Inheriting the symmetry of the simulation domain the absorbing region is a spherical shell with the external radius as the radius of the simulation domain ($R_{\rm sph}$) and the internal radius as $R_{\rm sph}-W_{\rm AB}$, where $W_{\rm AB}$ is the width of the absorbing spherical shell.
The parameters $\eta$ and $W_{\rm AB}$ are optimized for maximum absorption by the CAP to ensure that induced density higher than a cutoff value of $| 10^{-8}|\ e \rm\AA^{-3}$ is absorbed in the absorbing region during the propagation. 
\\

\begin{figure}
\includegraphics[width=.47\textwidth]{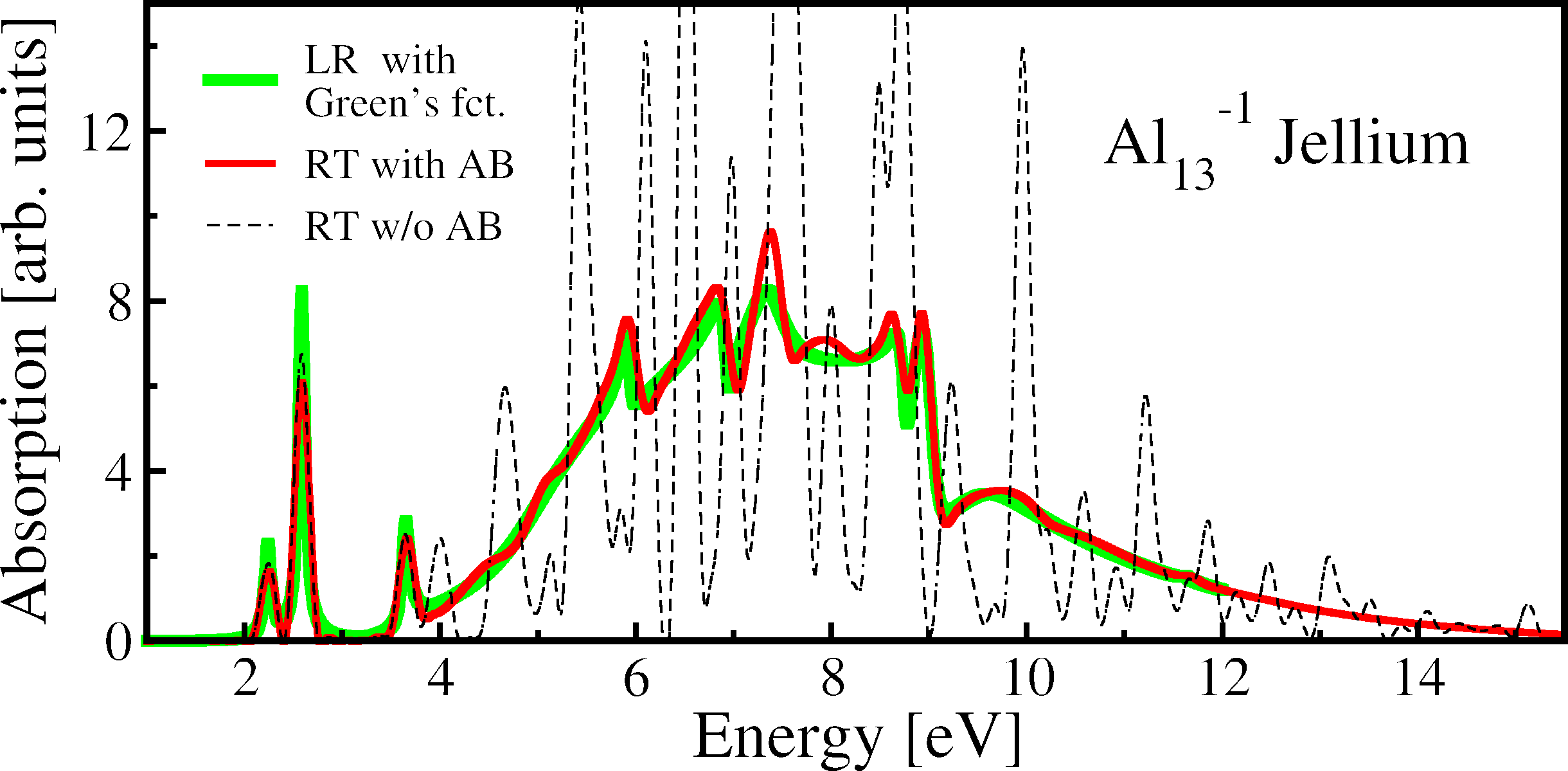}
\caption{
For Al$_{13}^-$ described by the spherical jellium model, a real-time TDDFT calculation using absorbing boundary in real-space is compared to the Green's function-based linear-response TDDFT calculation. The RT-TDDFT calculation without absorbing boundary is given for comparison, showing the impact of the coupling with the continuum.
}
\label{fig:jellium}
\end{figure}

To establish the reliability of using AB~\cite{DeGiovannini2015,octopusTancogne-Dejean2020} in a grid-based RT-TDDFT approach, we performed a comparison against a reference that can capture the contribution of the continuum to the absorption spectrum exactly. 
This reference is obtained by calculating the absorption spectrum using a Green's function-based linear-response (LR) TDDFT formalism~\cite{Lerme1998,Lerme2000} and the spherical jellium model where the atomistic structure is neglected. 
In this approach, owing to the spherical symmetry, the Green's function that contains all Kohn-Sham eigenfunctions, including the continuum, can be obtained analytically. Thus, this formalism using spherical jellium model provides an exact reference for the coupling to the continuum.
The details and references related to this approach are given in the Supplementary Materials. 

To compare with this reference, we carried out a grid-based RT-TDDFT calculation with absorbing boundary for the jellium model of Al$_{13}^-$. The two spectra shown in Figure~\ref{fig:jellium} are clearly identical. This confirms the reliability of real-time TDDFT with absorbing boundary conditions for calculating optical spectra when bound excited states lie within the ionization continuum. In addition, comparison with the spectra calculated without absorbing boundary underlines the huge importance of proper taking into account of the continuum.
~\\

\begin{figure}[h!]
\includegraphics[width=.48\textwidth]{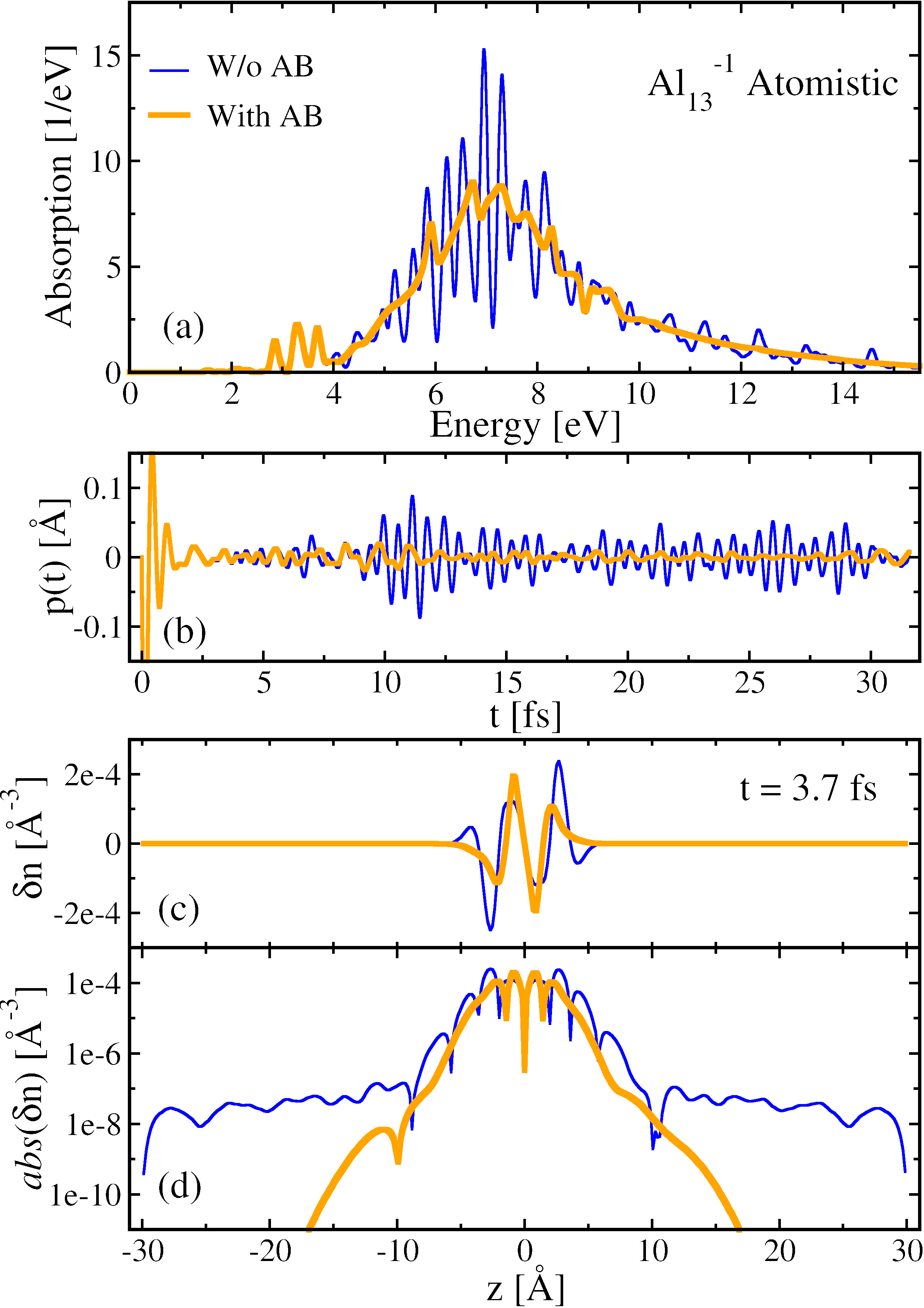}
\caption{
Comparison of (a) absorption spectra of Al$_{13}^-$ calculated in a spherical simulation domain of $R_{\rm sph}=30\,\rm\AA$ with (orange) and without (blue) the absorbing boundary, showing the strong influence of the coupling to the continuum. The corresponding time-dependent dipole moments (b) and snapshots of the induced densities (c,d) at $t=3.7$ fs underline the role of the reflected electron density. 
}
\label{fig:wNwoAB}
\end{figure}

Following this comparison on spherical jellium model, in Figure~\ref{fig:wNwoAB} we present grid-based RT-TDDFT results with (orange) and without (blue) absorbing boundary obtained on the actual system of interest, Al$_{13}^-$, described at the atomistic level. In panel (a), the spectra below 4\,eV correspond to bound-bound transitions and are identical between the two curves. It is immediately clear in these spectra that the use of the AB eliminates the unphysical peaks stemming from the reflections of excited electrons at the edges of the simulation box.  

The panel (c) shows a snapshot of the induced densities ($\delta n$) at time $t=3.7$ fs along the $z$-axis ($x=0,y=0$), $\hat{\mathbf{z}}$ being the direction of the perturbation.
The absolute value of the same $\delta n$ shown in a logarithmic scale in panel (d) demonstrates that induced density higher than a cutoff value of $10^{-8} e/\rm\AA^3$ is absorbed by the AB during the propagation.
This absorption of electron density represents the ionization mechanism during the time evolution and thus suppresses the unwanted reflections of excited electrons, which otherwise would hinder ionization and produce a different unphysical $\delta n$ (in blue).
The accumulation of this different $\delta n$ over the time spuriously interferes in the evolution of the dipole moment ($p(t)$) as witnessed by the differences captured in the time-dependent dipole moments shown in the panel (b).
The consequence of this difference appears in the spectra which are obtained from the Fourier transform of the respective $p(t)$. 
An animation showing the time evolution of the induced densities and depicting their consequence on the corresponding spectra is contained in the Supplemental Material.

Thus, Figure~\ref{fig:wNwoAB} demonstrates the importance, in real-time approaches, of properly accounting for the ionization process and ensuring that no artifacts are introduced, in order to correctly describe the system's dynamics and, consequently, the surface-plasmon resonance. The use of an absorbing boundary has allowed to clearly identify the broad spectral feature representing the SPR, far in the UV part ($4\,\mathrm{eV}\leqslant \mathrm{E} \leqslant 12\,\mathrm{eV}$) of the spectrum. Using AB we can now distinguish finer peaks that appear overlaid on this broad SPR. These features, having Fano-type~\cite{Fano1961} asymmetric line-shapes (e.g., at $\approx 6$ eV, or $8.7$ eV), are not distinguishable in the spectrum calculated without AB because of the interference arising due to the reflections.
\\

We have thus found an ab-initio method for calculating absorption spectra of metal clusters for energies above the IP by accurately including the contribution of the continuum within the  grid-based real-space real-time TDDFT. However, absorption spectra can also be obtained within the LR-TDDFT approach in terms of transition probability to excited states. In this approach, the numerical resolution is often done using expansion to localized basis sets (instead of using a spatial grid). Gaussian functions or other types of linear combination of atomic orbitals are the most common choices for the basis. However, they are not adequate to represent the continuum states which, at variance with bound discrete states, do not decay at large distances but rather display an oscillating behavior. One possible solution to this problem is the use of multicentric basis sets consisting of primitive functions that are products of radial spline functions and spherical harmonics~\cite{TiresiaCode_Decleva2022,TiresiaCode_Toffoli2024}. We performed calculations in which a large one-center expansion of such a B-spline basis set is supplemented by a few basis functions centered on the off-center nuclei, which guarantees fast convergence.  
The technical details of this calculation are given in the Supplemental Material.

\begin{figure}
\includegraphics[width=.47\textwidth]{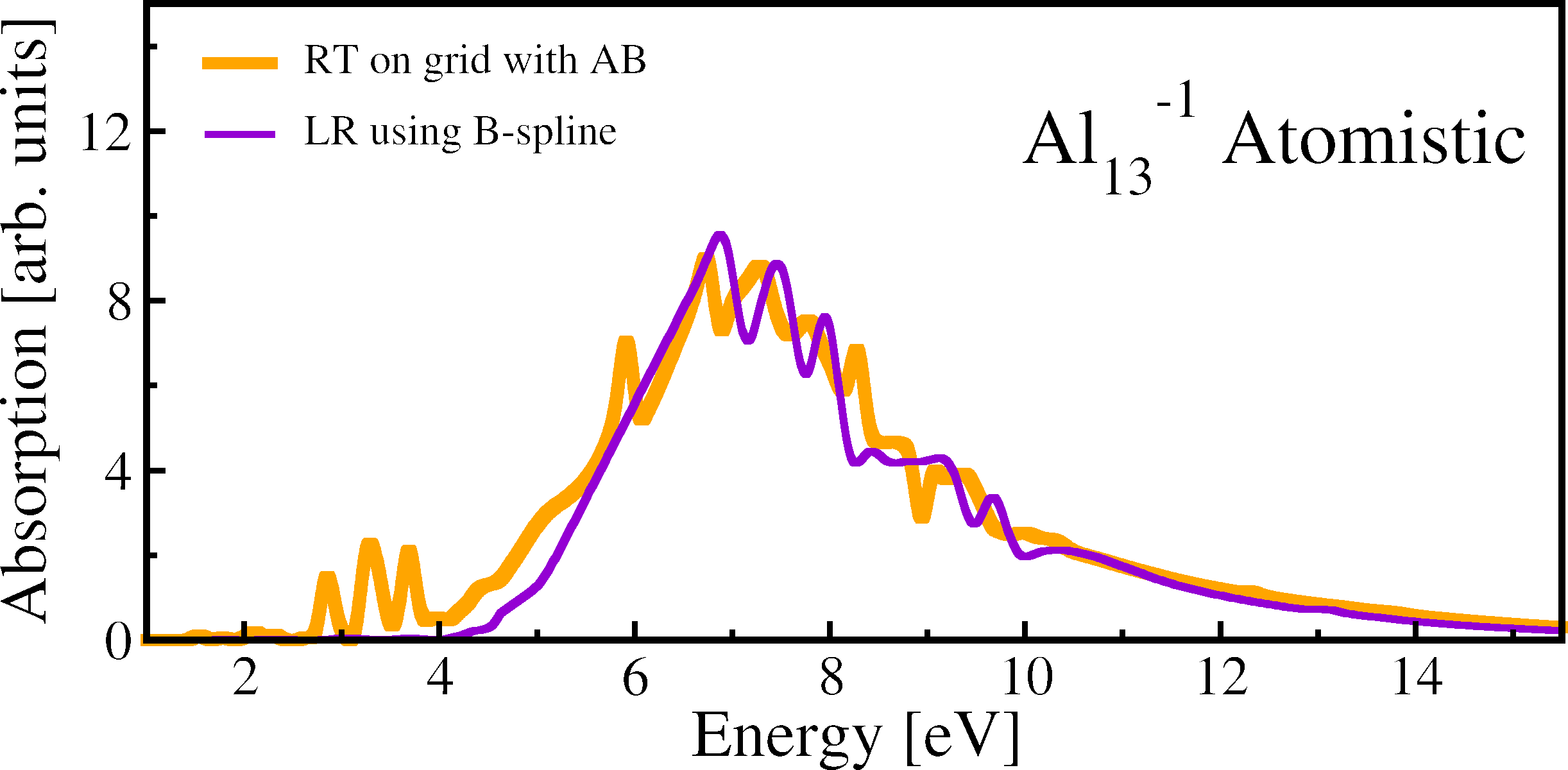}
\caption{
  Absorption spectrum of Al$_{13}^-$ calculated in RT-TDDFT using absorbing boundary in real-space compared to a LR-TDDFT calculation for energies above the IP using a multicenter B-spline expansion of the wave functions.
}
\label{fig:Bspline}
\end{figure}

The result of this alternative LR-TDDFT approach using multicentric B-splines is shown in Figure~\ref{fig:Bspline}. The absorption spectrum of Al$_{13}^{-}$ calculated in this approach for energies above the IP is compared with the spectra calculated earlier using RT-TDDFT with absorbing boundary. In spite of the very different approaches, the agreement is very good: the overall shape of the broad SPR and the finer structures are clearly obtained. This agreement demonstrates that once the essential factor, i.e., accurate consideration of the continuum, is taken into account, equivalent spectra are obtained in the basis-expansion based LR and grid-based RT approaches within TDDFT, regardless of the technical differences between them. At the same time, it resolves the existing discrepancy (cf., Supplemental Material) in the literature concerning the absorption spectrum of Al$_{13}^{-}$.
~\\

\begin{figure}
\includegraphics[width=.48\textwidth]{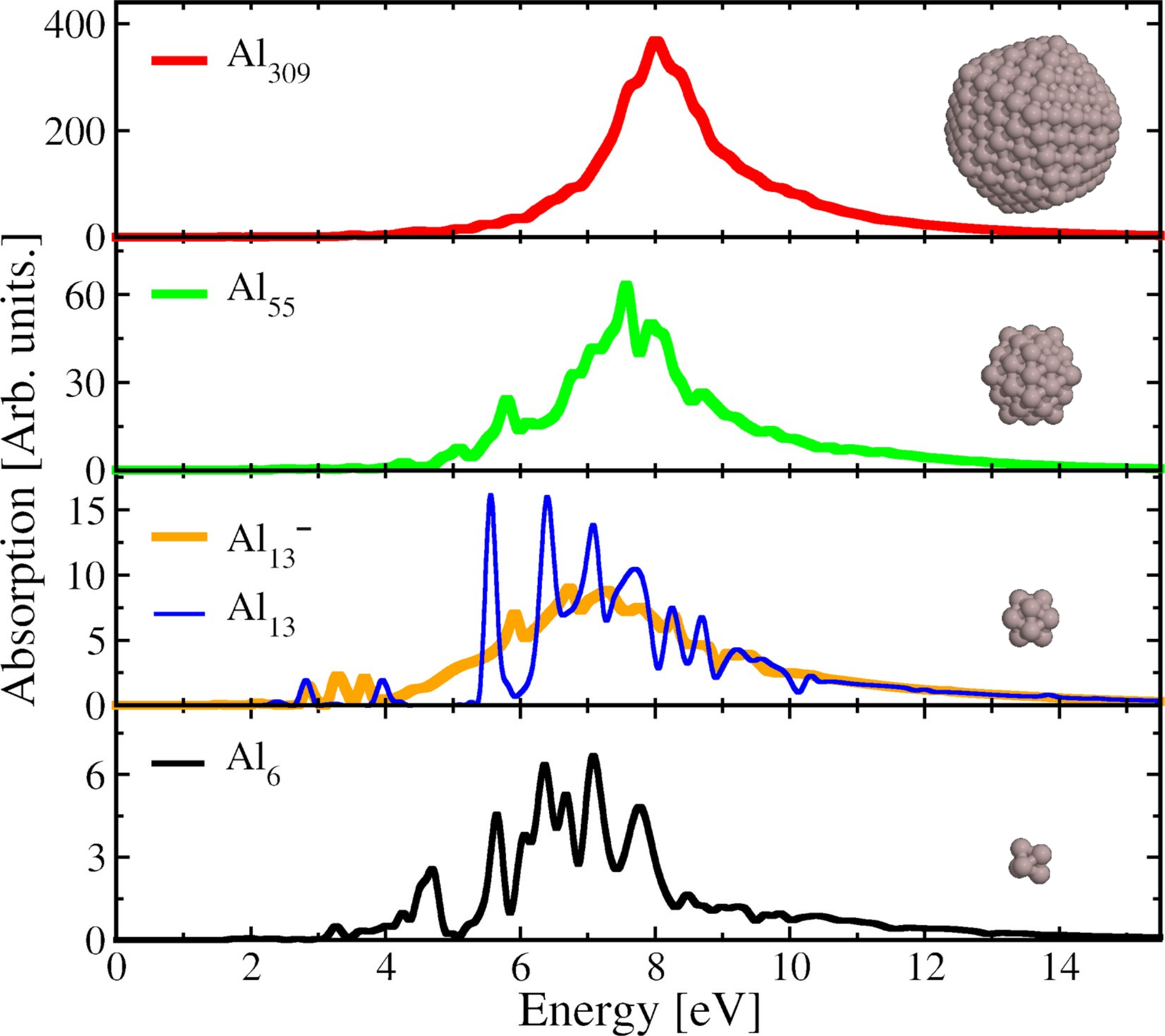}
\caption{
Evolution of the spectral features from discrete to plasmonic character as the number of atoms in the Al clusters grows from 6 to 309. In all the cases, the plasmon is significantly above the IP (the $E_{\rm HOMO}$ of the Kohn-Sham system is at -4.63\,eV (Al$_{309}$); -4.99\,eV (Al$_{55}$); -6.62\,eV (Al$_{13}$); -3.13\,eV (Al$_{13}^-$); and -6.75\,eV (Al$_{6}$)). 
}
\label{fig:plasmons-in-Al-clusters}
\end{figure}

As previously mentioned, a key advantage of the RT-TDDFT approach is its ability to calculate the spectra of large systems. Employing our rigorously validated method for simulating SPRs above the ionization potential, we calculate the absorption spectra of aluminum clusters of increasing size, shown in Figure~\ref{fig:plasmons-in-Al-clusters}. 

The spectra reveal the evolution of the excitation in the UV from molecular-like discrete features in Al$_{6}$ to a single broad SPR band in Al$_{309}$. Our calculations demonstrate the ability of RT-TDDFT with absorbing boundary to capture surface-plasmon resonances above the ionization potential in systems spanning a wide size range including clusters like Al$_{309}$ that are beyond the reach of LR-TDDFT approaches based on basis-set expansions.
\\

In conclusion, we have calculated the surface-plasmon resonances of aluminum clusters ``in the continuum,'' i.e., above the ionization threshold in the far UV. Doing so, we have shown the importance of properly taking into account the ionization process. This has been achieved by the methodical use of absorbing boundary conditions within the RT-TDDFT formalism in real space. The robustness of this approach is supported by obtaining the same result using Green's functions-based LR-TDDFT calculation on spherical jellium model which describes the contribution of the continuum exactly. The strength of the approach is highlighted by its ability to capture not only the surface-plasmon resonance as a broad spectral feature, but also the signatures of well-defined excited states in the continuum that appear as finer spectral features overlaid on the broad SPR. Without the proper inclusion of the ionization process, these fine structures cannot be distinguished because they are drowned in a multitude of perturbing spurious peaks.

Furthermore, we demonstrate, by comparing real-time and linear-response results, that linear-response TDDFT with a multicenter B-spline basis, an approach specifically developed for scattering problems, can accurately describe surface-plasmon resonances embedded in the ionization continuum.

Our calculations and the insights into the underlying mechanisms we have obtained resolve the discrepancies found in the literature for the optical spectra of small aluminum clusters. Our comparison of calculations using both RT- and LR-TDDFT approaches as well as different spatial representations (real-space grid and B-spline basis) is particularly instructive concerning the pitfalls of using TDDFT for calculating absorption above the ionization threshold of metal clusters.

Our results establish TDDFT as a reliable method able to describe spectra and localized surface-plasmon resonances at energies above the ionization threshold --- provided the effect to the ionization continuum is properly taken into account. The grid-based real-time formalism with absorbing boundary condition is particularly advantageous because it treats bound and continuum states on the same footing and it can be used to calculate high-energy spectra of large systems which are presently out of reach for linear-response calculations. The findings of the present study lay the ground work for the description of UV-plasmonic materials that are currently being studied in view of a multitude of applications that are unachievable with the conventional plasmonic materials.
~\\

\noindent\textit{Acknowledgments.} 
The authors thank Valérie Véniard, Pablo Gacía-González, and Matthias Hillenkamp for fruitful discussions. They acknowledge support from the French National Research Agency (Agence Nationale de Recherche, ANR) in the frame of the projects ``SchNAPSS,'' (ANR-21-CE09-0021) and ``GNOME'' (ANR-24-CE09-2403). M.C. thanks ED352 of Aix-Marseille University for the PhD scholarship. R.S.-R. acknowledges support from the Université Calude Bernard Lyon 1 (UCBL) in the frame of the programme ``Accueil Enseignants-Chercheurs'' AEC-2023, AEC-2024, and AEC-2025. Financial support from ICSC -- Centro Nazionale di Ricerca in High Performance Computing, Big Data and Quantum Computing, funded by European Union -- NextGenerationEU is gratefully acknowledged by M.S. and D.T. The work has used HPC resources from GENCI-IDRIS (Grant 2023-0906829, Grant A0170807662). Moreover, the authors acknowledge the contribution of the International Research Network IRN Nanoalloys (CNRS). 
~\\

R.S.-R. proposed this study, designed the RT-TDDFT simulations and conceived the idea of the manuscript. 
M.C., H.-C.W., and R.S.-R performed the grid-based RT-TDDFT simulations on spherical jellium model and ``atomistic'' clusters. 
J. L. performed the Green's function based LR-TDDFT calculation on spherical jellium model. 
D.T. and M.S. performed the B-spline LR-TDDFT calculation. 
R.S.-R. wrote the manuscript with F.R., H.-C.W., and V.D. and with input from all coauthors.
~\\

\noindent\textit{Data availability.}
All data presented in different plots within this paper and that related to the details of the calculation are available from the corresponding authors upon reasonable request.
~\\

\bibliography{refs_V7}

\end{document}